\documentclass[aps,pre,twocolumn,showpacs,superscriptaddress,10pt,longbibliography]{revtex4-1}
\usepackage[T1]{fontenc}

\pdfoutput=1
\usepackage[dvipsnames,rgb,dvips]{xcolor}
\usepackage{graphicx}
\usepackage{float}
\usepackage{overpic}
\usepackage{amsmath,amssymb,amsfonts}
\usepackage{hyperref}
\makeatletter
\def\amsbb{\use@mathgroup \M@U \symAMSb}
\makeatother
\usepackage{mathrsfs}

\newcommand\ve[1]{\boldsymbol{#1}}

\newcommand{\Reys}{\ensuremath{\textrm{Re}_{\rm s}}}

\newcommand{\ed}{\text{d}}
\newcommand{\ra}{\rightarrow}

\newcommand{\parti}[1]{\ensuremath{\partial_{#1}}}

\newcommand{\uouth}[2]{\ensuremath{\hat u^{(#1)}_{\text{o},#2}}}

\begin{document}
\title{Angular velocity of a {spheroid log rolling} in a simple shear at small Reynolds number}
\author{J. Meibohm}
\affiliation{Department of Physics, Gothenburg University, SE-41296 Gothenburg, Sweden}
\author{F. Candelier}
\affiliation{University of Aix-Marseille, CNRS, IUSTI UMR 7343, 13 013 Marseille, Cedex 13, France}
\author{T. Ros\'en}
\affiliation{KTH Mechanics, Royal Institute of Technology, SE-100 44 Stockholm, Sweden}
\author{J. Einarsson}
\affiliation{Department of Physics, Gothenburg University, SE-41296 Gothenburg, Sweden}
\author{F. Lundell}
\affiliation{KTH Mechanics, Royal Institute of Technology, SE-100 44 Stockholm, Sweden}
\author{B. Mehlig}
\affiliation{Department of Physics, Gothenburg University, SE-41296 Gothenburg, Sweden}

\begin{abstract}
We analyse the angular {velocity} of a small neutrally buoyant {spheroid log rolling} in a simple shear.  
When the effect of fluid inertia is negligible the {angular velocity $\ve \omega$ equals} half the fluid vorticity. 
We compute by singular perturbation theory how weak fluid inertia reduces the angular velocity {in an unbounded shear, and how this reduction depends upon the shape of the spheroid (on its aspect ratio). In addition we determine the angular velocity by direct numerical simulations. The results are in excellent agreement with the theory at small but not too small values of the shear Reynolds number, for all aspect ratios considered. For the special case of a sphere we find} $\omega/s = -1/2+0.0540\, \Reys^{3/2}$ where $s$ is the shear rate and $\Reys$ is the shear Reynolds number. This result differs from that derived by Lin {\em et al.} [J. Fluid Mech. {\bf 44} (1970) 1] who obtained a {numerical} coefficient roughly three times larger.
\end{abstract}
\maketitle

\section{Introduction}
{The angular dynamics of a small, neutrally buoyant particle in a viscous fluid is determined by the local fluid-velocity  gradients.
In the creeping-flow limit the particle rotates in such a way that the torque exerted by the fluid vanishes at every instant in time.}
For larger particles this is no longer true, in general. The particle angular momentum may change as the particle rotates, and the fluid near the particle is accelerated. It is usually a difficult problem to compute the effect of particle and fluid inertia on the dynamics of particles in flows. 

\begin{figure}[t]
\mbox{}\vspace*{1cm}
\begin{overpic}[width=4cm,clip]{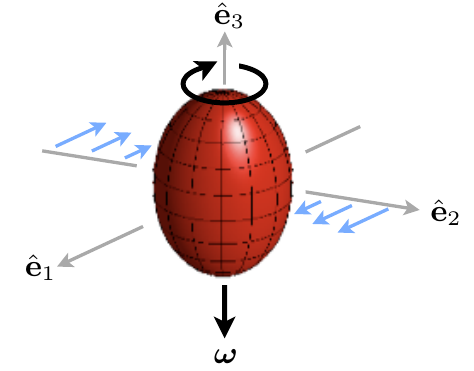}
\end{overpic}
\mbox{}\vspace*{-1cm}
\caption{\label{fig:1}
{Spheroid log rolling} in a simple shear. The $r_2$-coordinate is aligned with the shear direction, and the $r_3$-coordinate is the negative vorticity direction. {The axis of symmetry of the particle and its angular velocity $\ve \omega=\omega \ve{\hat e}_3$ are aligned with the flow vorticity.}
}
\end{figure}

In this paper we calculate the leading inertial correction to the angular velocity of a neutrally buoyant {spheroid that rotates in an unbounded shear
so that its axis of symmetry is aligned with the flow vorticity
(Fig.~\ref{fig:1}). This motion is referred to as \lq log rolling\rq{}. For an oblate spheroid this orbit is stable, whereas it is unstable for a prolate spheroid \cite{einarsson2015a}. The calculations summarised in this paper provide the corresponding angular velocity.}

{In the log-rolling orbit the locus of the particle surface does not change as a function of time. This simplifies the calculations considerably, because we can consider the steady problem. The latter is also easiest to analyse in direct numerical simulation because it is not necessary to change the numerical mesh as the particle moves.}

{In the log-rolling orbit the angular velocity aligns with the vorticity axis $\hat{\bf e}_3$ (Fig.~\ref{fig:1}), and
\begin{align}\label{eq:final_result} 
	\omega\equiv \ve \omega\cdot \hat{\bf e}_3  \sim -\frac{s}{2} + 0.0540 \,\frac{3sD}{10\pi}\,\Reys^{3/2}
\end{align}
to order $O(\Reys^{3/2})$.}
The first term on the r.h.s. corresponds to the angular velocity in the creeping-flow limit, equal to $-s/2$ for an unbounded shear with shear rate $s$ \cite{jeffery1922}. 
The second term on the r.h.s describes the inertial correction for small shear Reynolds number $\Reys \equiv a^2s/\nu$, and $\nu$ the kinematic viscosity of the fluid. 
{Furthermore $D$ is a shape parameter that depends on the aspect ratio of the spheroid. For a sphere we have $D=10\pi/3$, and the general expression is given in Eq.~(\ref{eq:d21}) below. It turns out that $D>0$ for all aspect ratios, so that} the effect of fluid inertia reduces the magnitude of the angular velocity. But we note that there are {other} examples where fluid inertia increases the angular velocity of a neutrally buoyant sphere \cite{blue2008}.

\begin{figure}[t]
\mbox{}\vspace*{1cm}
\begin{overpic}[height=7.cm,clip]{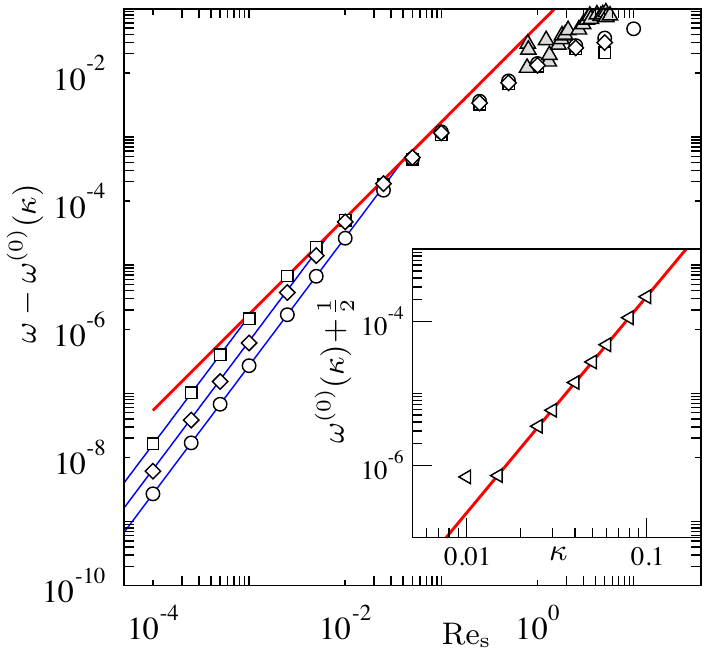}
\end{overpic}
\mbox{}\vspace*{-1cm}
\caption{\label{fig:2}
$\Reys^{3/2}$-correction to the angular velocity {of a spherical particle}.
{Open} symbols show results of {our} direct numerical simulations 
{(Section \ref{sec:dns})} for $\kappa=0.01$ ($\Box$), $0.025$ ($\Diamond$), and $0.05$ ($\circ$).
Here $\kappa=2a/L$, $a$ is the radius of the sphere, and $L$ is the linear dimension of the simulation domain. 
The thick solid (red) line shows Eq.~(\ref{eq:final_result}), 
{valid for an unbounded shear}.
The thin solid (blue) lines correspond to a quadratic $\Reys$-dependence, they show fits to Eq.~(\ref{eq:Re2}).  
Also shown are experimental results by Poe \& Acrivos ({Section \ref{sec:dc}}), filled triangles. The creeping-flow limit of the angular velocity in the finite, {bounded} system is denoted by $\omega^{(0)}(\kappa)$,
and $\omega^{(0)}(\kappa)\to -\tfrac{1}{2}$ as $\kappa\to 0$.
The inset shows
$\omega^{(0)}(\kappa){+\tfrac{1}{2}}$ as a function of $\kappa$ ($\triangleleft$). The thick solid line (red) {shows Eq.~(\ref{eq:wRe0}), with the fit-parameter
$C=0.22$.}}

\end{figure}

{For a spherical particle Eq.~\eqref{eq:final_result} reduces to $\omega/s\sim-1/2+0.0540\,\Reys^{3/2}$. This} is different from a result obtained by Lin {\em et al.} \cite{lin1970} who found a coefficient of $0.1538$ for the inertial correction, instead of $0.0540$. Our result is consistent with that of  Stone, Lovalenti \& Brady \cite{stone2016} who used a different technique. {We also performed direct numerical simulations of the problem to determine the angular velocity. For a spherical particle the results are shown in Fig.~\ref{fig:2}.  We see that Eq.~(\ref{eq:final_result}) agrees well with the simulation results when $\Reys$ is neither too large nor too small. }

Why do the numerical results agree with Eq.~(\ref{eq:final_result}) only at intermediate values of $\Reys$? This is a consequence of the fact that fluid inertia causes a singular perturbation of the creeping-flow problem. Even at very small shear Reynolds number the perturbation is not negligible far 
away from the particle, at distances
larger than the Saffman length $\ell_{\rm S}\equiv a /\sqrt{\Reys}$.
This means that a regular expansion of the solution to the particle (the \lq inner solution\rq{}) must match an approximate \lq outer solution\rq{}  that takes into account convective fluid inertia but does not fulfill the no-slip boundary conditions on the particle surface. 
The power $\Reys^{3/2}$ in  Eq.~(\ref{eq:final_result}) is due to the singular nature of the perturbation. The small parameter of the problem is $\epsilon \equiv \sqrt{\Reys}$.

The numerical simulations shown in Fig.~\ref{fig:2} were performed for a bounded shear, the simulation domain is a cube with side length $L$ (and we define $\kappa\equiv2a/L$). We expect
Eq.~(\ref{eq:final_result}) to agree with the numerical results when $L \gg \ell_{\rm S}$, in the opposite limit the perturbation is in effect regular. Fig.~\ref{fig:2} confirms this picture: the numerical results converge to Eq.~(\ref{eq:final_result}) as $L$ increases, but there are substantial deviations at small values of $\Reys$, as mentioned above. In this regime the inertial correction to the angular velocity is quadratic in $\Reys$ (thin solid blue lines in Fig.~\ref{fig:2}). That there is no term linear in $\Reys$ is a consequence of the symmetry of the problem. 
Naturally the perturbation theory must also fail at large $\Reys$. Eq.~(\ref{eq:final_result}) works reasonably well up to $\Reys\approx 5\times 10^{-2}$.

\begin{figure}[t]
\begin{overpic}[height=7cm]{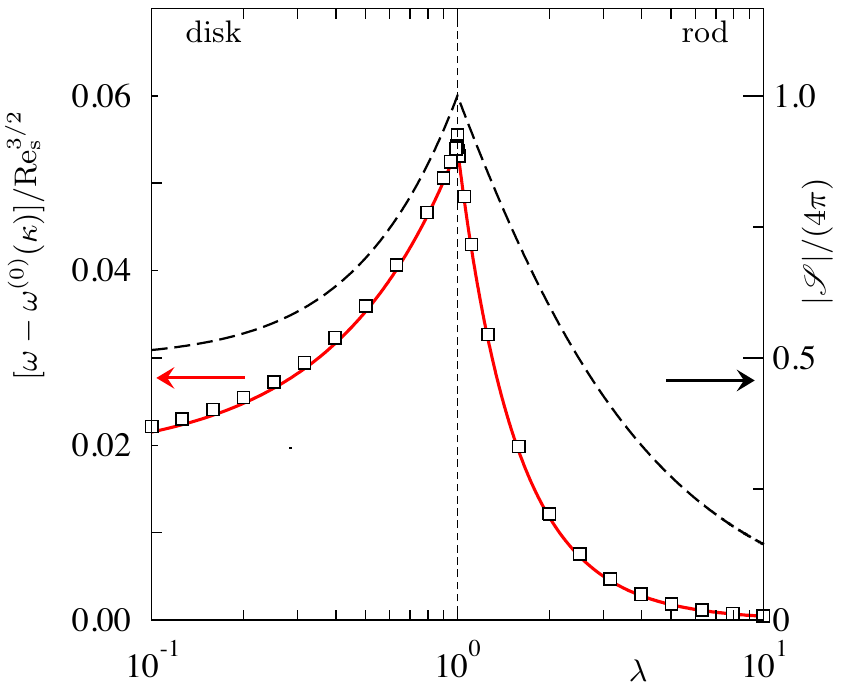}
\end{overpic}
\caption{\label{fig:3}
{$\Reys^{3/2}$-correction to the angular velocity of a neutrally buoyant spheroid log rolling in a simple shear, 
as a function of the aspect ratio $\lambda$. Shown are results of our direct numerical simulations (Section \ref{sec:dns}, $\Box$) for $\Reys=3 \times 10^{-3}$ and $\kappa = 0.01$. 
The theoretical result [Eq.~(\ref{eq:final_result}), valid for an unbounded shear] is shown as a solid red line. The red arrow points to the corresponding axis. Also shown is the surface area $|\mathscr{S}|$ of the particle (black dashed line). The corresponding axis is on the r.h.s. (black arrow).} }
\end{figure}

{For a spheroidal particle the steady-state angular velocity depends on the particle shape.
Fig.~\ref{fig:3} shows our results for the inertial correction of the angular velocity as a function of the particle aspect ratio at small but finite $\Reys$. 
For a prolate spheroid the aspect ratio is defined as $\lambda \equiv a/b$, where $a$ is the major semi-axis length of the particle, and $b$ is the minor semi-axis length. For oblate spheroids the aspect ratio is defined as $\lambda \equiv b/a$ \cite{einarsson2015a}.  Since the shear Reynolds number is defined in terms of the major semi-axis length of the particle we leave $a$  unchanged as we vary the particle shape.}

{In Fig.~\ref{fig:3},
the theoretical result (\ref{eq:final_result}) (red line) is compared to results of direct numerical simulations (white squares) of a large system ($\kappa = 0.01$) at $\Reys = 3 \times 10^{-3}$. We find excellent agreement.}

{The asymptotic matching method described in this paper is slightly different from that used by other authors, and has distinct advantages for the present problem. Therefore, we briefly comment on the differences here.}
{We obtain the outer solution as an expansion in $\ve k$-space, applying the approach described in Ref.~\cite{candelier2013:b}. But we use a new way of generating the terms in this expansion that  substantially simplifies the asymptotic matching.  
We obtain the angular velocity to order $\epsilon^3$ in terms of a lower-order solution, valid to order $\epsilon$.
The method employed by 
 Stone, Lovalenti \& Brady \cite{stone2016}, based on the reciprocal theorem, has the same} advantage.  Our calculations show that their method yields precisely the same integral expressions as our asymptotic matching. 
 It remains to be investigated how general this correspondence is.

\section{Formulation of the problem}
We consider a small neutrally buoyant 
{spheroid rotating in a simple shear with shear rate $s$. We assume that the axis of symmetry of the particle is aligned with the fluid vorticity (Fig.~\ref{fig:1}) and} that the centre-of-mass of the {particle} is advected by the flow.  
Our aim is to determine how weak {inertia affects} the angular dynamics. {In the log-rolling orbit angular velocity $\ve{\omega}$ aligns with the vorticity
axis $\hat{\bf e}_3$  at all times. }
We compute $\omega \equiv \ve \omega \cdot \hat{\bf e}_3$ 
as an expansion in $\epsilon \equiv \sqrt{\Reys}$:
\begin{align}
\omega = \omega^{(0)} + \epsilon \omega^{(1)} 
+ \epsilon^2 \omega^{(2)} + \epsilon^3 \omega^{(3)} +\ldots\,.
\end{align} 
We calculate {$\omega$} from the condition {that the torque must vanish in the steady state, $\ve \tau=0$.}
The torque is determined by the solution of Navier-Stokes equations:

\begin{align}\label{eq:nse}
\epsilon^2(\partial_tu_i\!+\!u_j\partial_ju_i) = -\parti{i} p+\partial_j\partial_j u_i\,,\,\,\,  \partial_j u_j=0\,,
\end{align}
written in dimensionless variables. As time scale
we take the inverse shear rate $s^{-1}$, as length scale we take the {length $a$ of the major axis of the particle}, and as velocity and pressure scales we take $as$ and $\mu s$, respectively. Here $\mu$ is the dynamic viscosity of the fluid. The torque that the fluid exerts upon the {particle} is measured in units of $ \mu s a^3$. All equations in the remainder of this article are written in dimensionless variables.

In Eq.~(\ref{eq:nse}), $u_i$ are the components of the fluid velocity and $p$ is {the} pressure. We use
the implicit summation convention that repeated indices are summed over from $1$ to $3$.  The boundary conditions are no slip on the surface {$\mathscr{S}$ } of the {particle}, and that the fluid velocity is undisturbed 
infinitely far away:
\begin{equation}\label{eq:bound}
        u_i = \varepsilon_{ijk} \omega_j r_k \,\, \text{for } {\ve r\!\in\!\mathscr{S}},\,\mbox{and}
\,\, u_i \ra u^{(\infty)}_i \,\, \text{as}\,\, r\ra\infty.
\end{equation}
Here $u_i^{(\infty)}$ are the components of the undisturbed fluid velocity, $r = |\ve r|$, $\ve r$ is the Cartesian coordinate vector with components
$(r_1,r_2,r_3)$, and the origin of the coordinate system is located at the centre of the {particle}.  Further $\varepsilon_{ijk}$ is the Levi-Civita symbol.
For the simple shear shown in Fig.~\ref{fig:1} we have
\begin{align}
\label{eq:simpleshear}
u_i^{(\infty)} =  A_{ij}^{(\infty)} r_j  \quad\mbox{with}\quad A_{ij}^{(\infty)} =\delta_{i1} \delta_{j2}\,.
\end{align}
Here $A_{ij}^{(\infty)}{\equiv \partial_j u^{(\infty)}_i}$ are the undisturbed fluid-velocity gradients, and $\delta_{ij}$ is the Kronecker symbol.

We decompose the fluid velocity into the undisturbed velocity $u_i^{(\infty)}$ and the disturbance
velocity $u_i'$, and decompose the pressure in a similar way:
\begin{align}\label{eq:dec}
        u_i     &=      u_i^{(\infty)} + u'_i\,,\quad  p       =      p^{(\infty)} + p'\,.
\end{align}
{Since the locus of the particle surface does not change as a function of time way may consider the steady disturbance problem:}
\begin{align}
        \epsilon^2(A^{(\infty)}_{jk} r_k \parti{j}u'_i\!+\!A^{(\infty)}_{ij}
u'_j\!+\!u'_j \parti{j}u'_i) \!=\! -\parti{i}p'\!+\!\partial_j\partial_j u'_i\,.
\label{eq:nse2}
\end{align}
The boundary conditions for $u'_i$ follow from Eq.~(\ref{eq:bound}):
\begin{subequations}
\label{eq:flowbound}
\begin{alignat}{3}
\label{eq:flowbound_inner}
u'_i    &=       \varepsilon_{ijk}\omega_j r_k- u^{(\infty)}_i  &\quad &\text{for } {\ve r\!\in\!\mathscr{S}}\,,        \\
u'_i&\to0       \quad   &	&\text{as }      r\ra\infty\,.
\label{eq:flowbound_outer}
\end{alignat}
\end{subequations}
Setting $\epsilon=0$ in Eq.~(\ref{eq:nse2}) corresponds to the Stokes limit. 
The disturbance flow in this limit, $u_i^{(0)}$, is well-known. 
At finite values of $\epsilon$, by contrast, Eq.~(\ref{eq:nse2}) is difficult to solve because the 
inertial terms on the l.h.s.~of Eq.~(\ref{eq:nse2}) constitute a singular perturbation of the Stokes problem.

{The singular nature of the perturbation is verified as follows.} The leading large-$r$ term in the Stokes solution of our problem
decays as $u_i^{(0)} \sim r^{-2}$. This allows us to estimate the terms on the l.h.s. of Eq.~(\ref{eq:nse2}) {as $r\to\infty$}:
\begin{subequations}
\begin{align}
\label{eq:8a}
\epsilon^2 A^{(\infty)}_{jk} r_k \parti{j} u_i^{(0)} &\sim  \epsilon^2  A^{(\infty)}_{ij} 
u_j^{(0)} \sim {\epsilon^2 } r^{-2}\,,\\
 \epsilon^2  u_j^{(0)} \parti{j} u_i^{(0)}   &\sim {\epsilon^2}r^{-5}\,.
\label{eq:8b}
\end{align}
\end{subequations}
The inertial terms (\ref{eq:8a}) balance the viscous term
\begin{align}
\partial_j\partial_j u_i^{(0)} \sim   r^{-4}
\end{align}
at  the Saffman length $\ell_{\rm S}$ (equal to $1/\epsilon$ in dimensionless variables).  Thus even at very small shear Reynolds number 
the effect of convective fluid inertia cannot be neglected at
distances larger than the Saffman length. We conclude that the problem
is singular, perturbation theory requires matched asymptotic expansions \cite{proudman1957,saffman1965}. In the \lq inner region\rq{} 
($r< \ell_{\rm S}$)  regular perturbation theory can be used. The inner solution satisfies the boundary condition (\ref{eq:flowbound_inner}) close to the particle. At $r \sim \ell_{\rm S}$ the inner solution is matched to a solution obtained in the \lq outer region\rq{} ($r>\ell_{\rm S}$) that takes into account convective fluid inertia and satisfies the boundary condition (\ref{eq:flowbound_outer}). 

\section{Method}
\subsection{Inner solution}
We seek regular perturbation expansions 
of the disturbance  velocity and the pressure in the inner region:
\begin{subequations}
\begin{align}
\label{eq:uinexpansion}
        {u}'_{\text{in},i}(\ve r) &= u^{(0)}_{{\rm in},i}(\ve r) + \epsilon \, u^{(1)}_{{\rm in},i}(\ve r) + \epsilon^2 \, u^{(2)}_{{\rm in},i}(\ve r) + 
\ldots \,,\\
{p}'_{\text{in}}(\ve r) &= p^{(0)}_{{\rm in}}(\ve r) + \epsilon \, p^
{(1)}_{{\rm in}}(\ve r) + \epsilon^2 \, p^{(2)}_{{\rm in}}(\ve r) +
\ldots \,.
\end{align}
\end{subequations}
The boundary conditions (\ref{eq:flowbound_inner}) read order by order:
\begin{subequations}
\begin{alignat}{2}
u^{(0)}_{{\rm in},i}(\ve r) &= {\varepsilon_{ijk}\omega_j^{(\infty)} r_k}- u^{(\infty)}_i  	\quad\,\quad\quad\mbox{for } {\ve r\!\in\!\mathscr{S}},\\
u^{(n)}_{{\rm in},i}(\ve r) &=  {\varepsilon_{ijk}\omega_j^{(n)} r_k} 		\quad\quad\mbox{for } n\geq 1 \mbox{ and } {\ve r\!\in\!\mathscr{S}}.
\end{alignat}
\end{subequations}
These boundary conditions do not suffice to determine the inner solution.
Therefore each term in the expansion (\ref{eq:uinexpansion}) is
matched to its counterpart in the outer solution at distance $r\sim \ell_{\rm S}$ from the particle.

\subsection{Outer solution} 
\label{sec:os}
To determine the outer solution we solve
\begin{align} \label{eq:nse3}
 \epsilon^2 (A^{(\infty)}_{jk} r_k \parti{j}u_{\text{o},i}'\!+\!A^{(\infty)}_{ij}u_{\text{o},j}' )  \!=\! -\parti{i}p'\!+\partial_j\partial_j u_{\text{o},i}'\!+ \!f_i
\end{align}
in the outer region. The subscript \lq o\rq{}  stands for \lq outer\rq{}.
In Eq.~(\ref{eq:nse3}) we have neglected the non-linear convective
term $u_j' \partial_j u_i'$. {At the end of this Section we verify that the error is of order $O(\epsilon^4)$.} 
The source term $f_i$ is introduced to account for the presence of the particle: $f_i({\ve r}) = D_{ij}\partial_j\,\delta({\ve r})$.
Here $\delta(\ve r)$ is the three-dimensional Dirac delta function. 
{The coefficients $D_{ij}$ must be determined so that the outer solution matches the inner solution. We expect that the disturbance flow due to the particle
can be represented in the outer region by a (symmetric) stresslet in the $\hat{\bf e}_1$-$\hat{\bf e}_2$ plane. Hence we make the {\em ansatz}
$D_{11}=D_{22}=D_{3i}=0$ and $D_{21}=D_{12}\equiv D$ (the \lq dipole strength\rq{} $D$ is determined by matching below).}

Since Eq.~(\ref{eq:nse3}) is linear it can be solved by Fourier transform.
In Section \ref{sec:outersol} we compute the expansion of 
the Fourier transform $\hat u_{{\rm o},i}'(\ve k)$ in powers of 
$\epsilon \equiv \sqrt{\Reys}$: 
\begin{align}
\label{eq:expT}
\hat u_{\text{o},i}'= \hat {\mathcal{T}_i}^{(0)} + \epsilon \hat {\mathcal{T}_i}^{(1)} +  \epsilon^2 \hat {\mathcal{T}_i}^{(2)} + \epsilon^3 \hat{\mathcal{T}_i}^{(3)} + \ldots\,.
\end{align}
It turns out that {the odd terms in} this expansion 
{are generalised functions of $\ve k$ \cite{candelier2013:b}. 
Transforming back to real space we find:}
\begin{subequations}
\label{eq:uoutr}
\begin{align}
\label{eq:T0}
{\mathcal{T}_i}^{(0)}(\ve r)&= - \frac{15D}{20\pi}
\frac{r_1 r_2 r_i}{r^5}\,,\\
\label{eq:T1}
{\mathcal{T}_i}^{(1)}(\ve r)&= 0\,,\\
\label{eq:T2}
{\mathcal{T}_i}^{(2)}(\ve r)&= - \frac{15 D}{240\pi}\Big\{\frac{1}{r^5}\big(r_1^2 r_2^2 r_i {-}r_1 r_2 r^2\varepsilon_{ij3}r_j\big) \\
\nonumber
& \hspace*{-2mm}\!{-}\!\frac{1}{r}\Big[\!-\!r_1\delta_{i1}\big(1\!-\!\frac{r_1^2}{3r^2}\big)
\!+\!\delta_{i2}r_2\big(1\!+\!\frac{r_2^2}{3r^2}\big){-\!\delta_{i3}\frac{r_3^3}{3r^2}}
\Big] \Big\}
\,,\\
\label{eq:T3}
{\mathcal{T}_i}^{(3)}(\ve r) &= \frac{3 D}{10\pi}A'_{ij} r_j \,.
\end{align}
\end{subequations}
{Details are given in  Section \ref{sec:outersol}. Eqs.~}\eqref{eq:uoutr} can be used to {verify}  that the non-linear convective term $u_j' \partial_j u_i'$ can be neglected in the outer problem \eqref{eq:nse3} to order $\epsilon^3$. 
The argument goes as follows. Eq.~\eqref{eq:uoutr} shows that ${\mathcal{T}_i}^{(n)}(\ve r)$ scale as $\sim r^{n-2}$ for $n=0,\ldots,3$ in the matching region. In this region, $r$ is of order of $\epsilon^{-1}$. Using \eqref{eq:expT} it follows that all terms in Eq.~\eqref{eq:nse3} are of order $\mathcal{O}(\epsilon^4)$ in the matching region:
 \begin{equation}
\epsilon^2 \big[A^{(\infty)}_{jk} r_k \parti{j}{\mathcal{T}_i}^{(n)} \!+\!A^{(\infty)}_{ij}
 {\mathcal{T}_j}^{(n)}  \big] \!\sim \!\partial_j \partial_j {\mathcal{T}_i}^{(n)} \!\sim\! O(\epsilon^4)
 \end{equation}
 for all $n=0,\ldots,3$. Using the same arguments we can also estimate the 
magnitude of the neglected non-linear convective term in the matching region:
\begin{align}
\epsilon^2 {\mathcal{T}_j}^{(n)} \partial_j {\mathcal{T}_i}^{(n)} \sim O(\epsilon^7)\,.
\end{align}
It turns out that the non-linear convective term remains negligible for $n=0,\ldots,3$ when $r > \ell_{\rm S}$.
We conclude that the non-linear convective
term can be neglected in the outer problem, because it is 
sub-leading within and beyond the matching region for $n=0,\ldots,3$. 
In the inner problem, in general, the non-linear convective term cannot be neglected for $n\leq 3$. 

\subsection{Matching}
\label{sec:match}
The outer solution (\ref{eq:uoutr}) provides boundary conditions
for the inner problem in the matching region, at $r \sim 1/\epsilon$.  
The leading-order terms of $u'_{{\rm in},i}$ at large $r$ must match
Eq.~(\ref{eq:uoutr}).
We now discuss the matching order by order.
To leading order $\epsilon^0$ the inner problem reads:
\begin{align}
 -\parti{i} p_{\rm in}^{(0)}\!+\!\partial_j\partial_j 
u_{{\rm in},i}^{(0)}= 0\,.
\end{align}
The boundary and matching conditions are given by:
\begin{subequations}\label{eq:flowbound_inner_0}
\begin{alignat}{3}
u_{{\rm in},i}^{(0)}	&= \varepsilon_{ijk}\omega_j^{(0)} r_k- u^{(\infty)}_i &	\quad &\text{for } {\ve r\!\in\!\mathscr{S}}\,,\\
u_{{\rm in},i}^{(0)} &\sim  {\mathcal{T}_i}^{(0)}	&	&\text{as }r\to \infty\,. 
\label{eq:m0}
\end{alignat}
\end{subequations}
{From Eq.~(\ref{eq:m0}) we see that ${\mathcal{T}_i}^{(0)}$ is the leading-order contribution to the Stokes problem of a spheroid freely rotating} in a simple shear {with its symmetry axis aligned with the $\hat{\bf e}_3$-axis. The matching condition \eqref{eq:m0} determines the dipole strength $D$ introduced in the previous Section. One finds $D$ as a function of the aspect ratio $\lambda$:
\begin{subequations}\label{eq:d21}
\begin{equation}
\label{eq:d21p}
D\! =\!  \frac{16\pi(\lambda^2-1)^3}{3\lambda^3[5\lambda -7\lambda^3+2\lambda^5+ 3\sqrt{\lambda^2-1}
\,\text{arccosh}(\lambda)]}
\end{equation}
for a prolate spheroid ($\lambda >1)$, and
\begin{equation}
\label{eq:d21o}
D =  -\frac{16\pi(1-\lambda^2)^3}{3[5\lambda -7\lambda^3+2\lambda^5- 3\sqrt{1-\lambda^2}\,\text{arccos}(\lambda)]}
\end{equation}
\end{subequations}
for an oblate spheroid ($\lambda < 1$).}  The corresponding angular velocity is {the Jeffery-result} $\omega^{(0)} = -\tfrac{1}{2}$ \cite{jeffery1922}.

The {order-$\epsilon^1$} problem reads:
\begin{subequations}
\begin{align}   
 -&\parti{i}p_{\rm in}^{(1)}\!+\!\partial_j\partial_j u_{{\rm in},i}^{(1)}= 0\,,\\
&u_{{\rm in},i}^{(1)}= {\varepsilon_{ijk}\omega_j^{(1)} r_k}\,\,  \text{for}\,\, {\ve r\!\in\!\mathscr{S}}\,,
\end{align}
together with the matching condition
\begin{align}	
\quad u_{{\rm in},i}^{(1)} \sim \mathcal{T}^{(1)}_i \quad \text{as }   r \to \infty\,.
\end{align}
\end{subequations}
Since  $\mathcal{T}^{(1)}_i$ vanishes [see Eq.~\eqref{eq:T1}] 
there is no term in Eq.~\eqref{eq:expT} to which the flow $u_{{\rm in},i}^{(1)}$ produced by the rotating particle can be matched.  {We conclude that $u_{{\rm in},i}^{(1)}\equiv0$. It follows that} the $\epsilon^1$-contribution
to the angular velocity must vanish, $\omega^{(1)}=0$.

The second-order problem is inhomogeneous:
\begin{subequations}
\begin{align}
\nonumber
 &-\parti{i}p_{{\rm in}}^{(2)}\!+\!\partial_j\partial_j u_{{\rm in},i}^{(2)}\\
 &\hspace*{5mm}=   A^{(\infty)}_{jk} r_k \parti{j}u_{{\rm in},i}^{(0)}\!+\!A^{(\infty)}_{ij} u_{{\rm in},i}^{(0)} +\!{u_{{\rm in},j}^{(0)}} \parti{j}u_{{\rm in},i}^{(0)}\,,
\label{eq_uprime2}
\end{align}
with boundary and matching conditions:
\begin{alignat}{4}\label{eq:bound_match_2}
u_{{\rm in},i}^{(2)}&={\varepsilon_{ijk}\omega_j^{(2)} r_k} &\quad	&\text{for } {\ve r\!\in\!\mathscr{S}}\,,	\\
u_{{\rm in},i}^{(2)} &\sim {\mathcal{T}_i}^{(2)} 	&	&{\text{as } r \to \infty\,.}\label{eq:bound_match_2c}
\end{alignat}
\end{subequations}
The solution is the sum of a homogeneous and a particular part:
$u_{{\rm in},i}^{(2)} = u_{{\rm h},i}^{(2)} + u_{{\rm p},i}^{(2)}$.
The particular solution is found to be 
$u_{{\rm p},i}^{(2)} = {\mathcal{T}_i}^{(2)}$ +{{${O}(1/r^2)$}}.
The homogeneous part solves the Stokes problem
\begin{subequations}
\begin{align}
 -\parti{i}{p}^{(2)}_{\rm h}\!+\!\partial_j\partial_j  u_{{\rm h},i}^{(2)} =  0
\end{align}
with boundary conditions
\begin{align}
\label{eq:bc5}
u_{{\rm h},i}^{(2)}	&=	
\varepsilon_{ijk}\omega_j^{(2)} r_k  - {u_{{\rm p},i}^{(2)}} \,\, \text{for } {\ve r\!\in\!\mathscr{S}}\,.
\end{align}
\end{subequations}
{The explicit solution of the homogeneous problem shows that
$u_{{\rm h},i}^{(2)}$  is asymptotic to 
${O}(1/r^2)$ as $r\to\infty$.} {This implies that the matching condition \eqref{eq:bound_match_2c} is fulfilled by $u_{\text{p},i}^{(2)}$ alone.}

Direct computation confirms that
the $\epsilon^2$-contribution to the angular velocity of the particle must vanish {for torque-free rotation}, $\omega^{(2)}=0$.
We remark {that there are}, however, examples where there is a linear inertial correction to the angular velocity \cite{candelier2016}.

{So far we have found that $\omega^{(0)} = -\tfrac{1}{2}$, and
$\omega^{(1)} = \omega^{(2)}=0$. Now consider the third order
in $\epsilon$.  At this} order the problem is homogeneous since 
$u_{{\rm in},i}^{(1)} \equiv 0$.
The equations to solve are
\begin{subequations}
\label{eq:thirdorder}
\begin{align}   
 -&\parti{i}p_{\rm in}^{(3)}\!+\!\partial_j\partial_j  u_{{\rm in},i}^{(3)}=  0\,,\\
&u_{{\rm in},i}^{(3)}= \varepsilon_{ijk}\omega^{(3)}_j r_k \quad  \text{for } {\ve r\!\in\!\mathscr{S}}\,,
\end{align}
together with the matching condition
\begin{align}\label{eq:matching_3}
u_{{\rm in},i}^{(3)}  \sim {\mathcal{T}_i}^{(3)}  \quad {\text{as } r \to\infty\,.}
\end{align}
\end{subequations}
Eq.~\eqref{eq:T3} shows that $\mathcal{T}^{(3)}_i$ is a linear flow. 
The third-order problem (\ref{eq:thirdorder}) is thus equivalent to that of a freely rotating {speroid} in the linear flow ${ A}'_{ij} r_j$ as $\epsilon\to0$.
The solution of this problem is known, {it is discussed in the next
Section.}

\subsection{Angular velocity} 
\label{sec:av}
The third-order correction to the
angular velocity is computed by requiring that the third-order
torque $\ve \tau^{(3)}\equiv \tau^{(3)} \,\hat{\bf e}_3$ vanishes. {The third-order inner problem is equivalent to the problem of determining the torque on a freely rotating spheroid in a linear flow in the creeping-flow limit. The solution is known (see e.g. p.64 in \cite{kim1991}). In the log-rolling orbit we have:}
\begin{align}
\label{eq:tau3}
\tau^{(3)} = 8 \pi {\big[  \tfrac{3D}{20\pi}(A'_{21}-A'_{12})-\omega^{(3)}\big]}\,.
\end{align}
Setting Eq.~(\ref{eq:tau3}) to zero yields $\omega^{(3)}$, and thus:
\begin{align}
\label{eq:theory}
{\omega \sim -  \tfrac{1}{2} + \tfrac{3 D}{20\pi} (A'_{21}-A'_{12}) \epsilon^3}
\end{align}
in dimensionless variables. In Section \ref{sec:outersol} we show how to evaluate the constant
coefficients $A'_{ij}$. The result is
\begin{align}
A'_{12} &= 0.0328\,,\quad A'_{21} = 0.1408 \,.
\end{align}
{It follows that}
\begin{align}\label{eq:result}
        \omega&{\sim} -\tfrac{1}{2} + 0.0540 \,\frac{3D}{10\pi}\,\epsilon^3\,.
\end{align}
In dimension{al} variables this result corresponds to Eq.~(\ref{eq:final_result}).

\section{Solution of the outer problem}\label{sec:outersol}
In this Section we determine the solution of the outer problem (\ref{eq:nse3}) {in the limit $\epsilon\to0$}. The results are Eqs.~(\ref{eq:uoutr}).

Since Eq.~(\ref{eq:nse3}) is linear it can be solved by Fourier transform
(we employ the symmetric convention).
Using incompressibility to eliminate the pressure,
we find for the Fourier transform  $\hat u'_{\text{o},i}({\ve k})$:
\begin{align}\label{eq:outshear}
\epsilon^2 & \left[k_1\frac{\partial \hat u'_{\text{o},i}}{\partial k_2}
 -\Big(\delta_{1i}-\frac{2  k_1 k_i}{ k^2}  \Big)\hat u'_{\text{o},2}\right] \\
&\hspace*{2mm}={{k^2} \hat u'_{\text{o},i} + \frac{{\rm i}D}{(2\pi)^{3/2}  k^2}\left(2 k_i  k_1  k_2-  k^2(k_1\delta_{i2}+k_2\delta_{i1})\right)\,.}
\nonumber
\end{align}
To determine the terms ${\cal T}^{(n)}_i$ in Eq.~(\ref{eq:expT}) we expand $\hat u'_{\text{o},i}$ 
in $\epsilon$. {We see in the next Section, however, that} a regular expansion does not give all required terms.
There are terms involving generalised functions that must be determined
separately (Section \ref{sec:singexp1}).

\subsection{{Even-order terms}}
\label{sec:regexp}
It follows from Eq.~(\ref{eq:outshear}) that the leading order in the $\epsilon$-expansion of $\hat u'_{{\rm o},2}$ takes the form:
\begin{align}
\label{eq:u1out2}
\hat u'_{\text{o},2} &
= -{\frac{{\rm i}D}{(2\pi)^{3/2}}\frac{k_1}{k^4}\left(2 k_2^2-k^2\right)        +O( \epsilon) \,.}
\end{align}
{This determines $  \mathcal{\hat T}_2^{(0)}({\ve k})$:}
\begin{subequations}
\begin{align}
        \mathcal{\hat T}_2^{(0)}({\ve k}) ={ \frac{{\rm i}D}{(2\pi)^{3/2}}\frac{k_1}{k^4}\left(k^2-2 k_2^2\right) }\,.
\end{align}
This is the Fourier transform of Eq.~(\ref{eq:T0}). The next term in the expansion reads:
\begin{align}\label{eq:T2n12}
       \mathcal{\hat T}_2^{(2)}({\ve k}) &={ -\frac{4 \,{\rm i}D}{(2\pi)^{3/2}}\frac{k_1^2 k_2 }{k^8}\left(k^2-k_2^2\right)\,.}	
\end{align}
This is the Fourier transform of Eq.~(\ref{eq:T2}). The next order {in the expansion} is:
\begin{align}
       \mathcal{\hat T}_2^{(4)}(\ve k) &= -\frac{4 \,{\rm i}D}{(2\pi)^{3/2}}\frac{k_1^3}{k^{12}}\left(k^2-6 k_2^2\right)\,.
\end{align}
\end{subequations}
In a similar way we expand $u'_{\text{o},1}$. This yields
\begin{subequations}
\begin{align}
        \mathcal{\hat T}_1^{(0)}(\ve k) &\!=\! {\frac{{\rm i}D}{(2\pi)^{3/2}}\frac{k_2}{k^4}( k^2-2 k_1^2)\,,  }\\
        \mathcal{\hat T}_1^{(2)}(\ve k) &\!=\! {\frac{4\,{\rm i}D}{(2\pi)^{3/2}}\frac{k^3_1k_2^2}{k^8}\,,    }\\
        \mathcal{\hat T}_1^{(4)}(\ve k) &\!=\! {\frac{4\,{\rm i}D}{(2\pi)^{3/2}}\frac{k_1^2k_2}{k^{12}}\big[k_1^4\!+\!(2 k_3^2-5 k_2^2) k_1^2\!+\!k_3^2
(k_2^2\!+\!k_3^2)\big]\,.}
\end{align}
\end{subequations}
We see that a regular expansion of Eq.~(\ref{eq:u1out2})
yields only even orders ${\cal T}^{(2n)}_i$. Yet 
odd orders  ${\cal T}^{(2n-1)}_i$
are required to match the inner solution.
Terms corresponding to odd powers in $\epsilon$ {are computed} in the following Sections.\\[1ex]

\subsection{{Odd-order terms}}\label{sec:singexp1}
Odd terms in the expansion (\ref{eq:expT}) are not
captured by the  expansion in $\epsilon$ {described in the previous Section} because the odd-order terms are zero everywhere except at isolated points in $\ve k$-space.
{To compute the odd-order terms we}  subtract the lowest term(s) in the expansion (\ref{eq:expT}) to make the next {odd}  order leading. This new leading order is then found by taking {the limit $\epsilon\to0$}. This idea was pioneered by \citet{childress1964} and \citet{saffman1965}, and employed by \citet{lin1970}. 
{These papers start from an ansatz of the outer solution written in terms of a \lq stretched\rq{} configuration-space variable $\tilde{\ve r} \equiv \epsilon\ve r$, see for example Eq.~(3.12) in Ref.~\cite{saffman1965}. 
}
Here we pursue an alternative path that {does not absorb the $\epsilon$ into a new variable $\ve{\tilde r}$ and} proceeds with generalised functions in $\ve k$-space 
\cite{candelier2013:b}. We define:
\begin{align}\label{eq:u1ni}
        \hat u^{(q)}_{\text{o},i}({{\ve k}}) &\equiv \frac1{\epsilon^{q-1}}
\Big[\hat u'_{\text{o},i}({ \ve k})
- \sum_{k=0}^{q-1}\epsilon^k\mathcal{\hat T}^{(k)}_i({\ve k})\Big]\\
&\sim \epsilon\,\mathcal{\hat T}^{(q)}_i(\ve k) + \epsilon^2 \mathcal{\hat T}^{(q+1)}_i(\ve k) + \ldots\,.\nonumber
\end{align}
To extract an odd term corresponding
to $q=2n-1$ {one can} take the limit  $\epsilon \to 0$ \cite{candelier2013:b}:
{\begin{subequations}\label{eq:Tepslim}
\begin{align}\label{eq:Tepslima}
	\mathcal{\hat T}^{(2n-1)}_i(\ve k)&=
        \lim_{\epsilon\to0}\left[\epsilon^{-1}\hat u^{(2n-1)}_{\text{o},i}(\ve k)\right]\,.
\end{align}
{It turns out, however, that an alternative procedure has distinct advantages:}
\begin{align}\label{eq:Tepslimb}
	\mathcal{\hat T}^{(2n-1)}_i(\ve k)&=
        \lim_{\epsilon\to0}\frac{\rm d}{\rm d \epsilon}\left[\epsilon^{-1}\hat u^{(2n-2)}_{\text{o},i}(\ve k)\right]\,.
\end{align}
\end{subequations}}
Eqs.~(\ref{eq:Tepslim}) are evaluated {as follows}. We derive a differential equation for  $\hat u^{(2n-1)}_{\text{o},i}{(\ve k)}$ by substituting Eq.~(\ref{eq:u1ni}) into Eq.~\eqref{eq:outshear}:
\begin{align}\label{eq:u2n}
&\epsilon^2 k_1\frac{\partial \hat u^{(2n-1)}_{\text{o},i}}{\partial k_2}\!-\!k^2 \hat u^{(2n-1)}_{\text{o},i} \!-\! \epsilon^2\Big(\delta_{1i}\!-\!\frac{2 k_1 k_i}{k^2}  \Big)\hat u^{(2n-1)}_{\text{o},2} \\
&\hspace*{5mm}=
- k^2 \mathcal{\hat T}^{(2n)}_i\,.\nonumber
\end{align}
For $i=1,2$ the solutions of  Eq.~\eqref{eq:u2n} are:
\begin{widetext}
\begin{subequations}\label{eq:u12nifin}
\begin{align}\label{eq:u12n2fin}
        \hat u^{(2n-1)}_{\text{o},2}(\ve k)
&= -\frac{1}{k^2}\int_0^\infty\!\!\!\!
\ed \xi\,  {\rm e}^{-{\xi}\big({k_1^2}\xi^2/3+k_1k_2\xi + k^2\big)/\epsilon^2}
\Big[k^4\mathcal{\hat T}^{(2n)}_2(\ve k)\Big]_{k_2=k_1\xi+k_2}\,,
\\
                \hat u^{(2n-1)}_{\text{o},1}(\ve k)   &= -\int_0^\infty \!\!\!\!\ed \zeta\,
{\rm e}^{-{\zeta} \big({k_1^2}\zeta^2/3+k_1k_2\zeta + k^2\big)/\epsilon^2} \Big[k^2\mathcal{\hat T}^{(2n)}_1(\ve k) +\frac{k^2-2k^2_1}{k_1k^2}\hat u^{(2n-1)}_{\text{o},2}(\ve k) \Big]_{k_2=k_1\zeta+k_2}  \,,
\end{align}
\end{subequations}
\end{widetext}
Then, to determine $\mathcal{\hat T}^{(2n-1)}_i$ we must take the limit $\epsilon\to 0$ in Eqs.~(\ref{eq:Tepslim}). To this end
we use the fact
that $\hat u^{(2n-1)}_{\text{o},i}(\ve k)$ are integrable functions
of $\ve k$. For such a function the limit $\epsilon\to0$ can be expressed
in terms of generalised functions. We need the following three relations. First,
\begin{subequations}\label{eq:deltaform} 
\begin{equation}\label{eq:deltaform1} 
\lim_{\epsilon\to0}\frac1{\epsilon^{3}}f(\ve k/\epsilon) 
= \mathcal{A}\, \delta(\ve k) \quad\mbox{with}\quad
 \mathcal{A}= \int_{\mathbb{R}^3} \!\!\!\!\ed^3k\,f(\ve k)\,.
\end{equation}
Second, consider a function $g$ that integrates to zero, 
\begin{equation}
\int_{\mathbb{R}^3}  \ed^3k\,g(\ve k)=0\,.
\end{equation}
In this case we use
\begin{equation}\label{eq:deltaform2}
\hspace*{-4mm}\lim_{\epsilon\to0}\frac{g(\ve k/\epsilon)}{\epsilon^{4}}
\!=\! \mathcal{A}_{i}\frac{\partial \delta(\ve k)}{\partial{k_{i}}}
\,\,\mbox{with}\,\,
        \mathcal{A}_{i}\!=\!-\int_{\mathbb{R}^3} \!\!\!\!\ed^3k\,k_{i}
g(\ve k)\,.
\end{equation}
{Third, we make use of the identity:}
\begin{equation}\label{eq:deltaform3}
\hspace*{-3mm} \lim_{\epsilon\to0}\frac{\rm d}{\rm d \epsilon}\left[\frac{h(\ve k/\epsilon)}{\epsilon^{3}}\right]
\!=\! \mathcal{A}_{i}\frac{\partial \delta(\ve k)}{\partial{k_{i}}}
\,\,\mbox{with}\,\,
        \mathcal{A}_{i}\!=\!-\int_{\mathbb{R}^3} \!\!\!\!\ed^3k\,k_{i}
h(\ve k)\,.
\end{equation}
\end{subequations}
{This relation follows from Eq.~\eqref{eq:deltaform1} for a differentiable function $h(\ve k)$ (Appendix \ref{app:delta}).}
To make use of Eqs.~(\ref{eq:deltaform}) we
first derive a scaling relation for $ \mathcal{\hat T}_i^{(2n)}$ applying the results of Section \ref{sec:regexp}:
\begin{align}
 \mathcal{\hat T}_i^{(2n)}(\ve k) &= \frac1{\epsilon^{2n{+1}}}\mathcal{\hat T}_i^{(2n)}(\ve k/\epsilon)\,.
\label{eq:b}
\end{align}
This relation together with \eqref{eq:u12nifin} implies:
\begin{align}\label{eq:u12nscal}
        \hat u^{(2n-1)}_{\text{o},i}(\ve k)   &=      \frac1{\epsilon^{2n-1}} \hat u^{(2n-1)}_{\text{o},i}(\ve k/\epsilon)\,.
\end{align}
Eq.~(\ref{eq:u12nscal}) allows us to bring (\ref{eq:Tepslim})  into
a form where Eqs.~(\ref{eq:deltaform}) can be directly applied. {This yields:}
\begin{subequations}\label{eq:tnlimit}
\begin{align}\label{eq:tnlimit1}
\mathcal{\hat T}^{(2n-1)}_i(\ve k)
&= \lim_{\epsilon\to 0} \frac{1}{\epsilon^{2n}}
\hat u^{(2n-1)}_{\text{o},i}(\ve k/\epsilon)\,,	\\
\mathcal{\hat T}^{(2n-1)}_i(\ve k)
&= \lim_{\epsilon\to 0} \frac{\rm d}{\rm d \epsilon}\left[\frac{1}{\epsilon^{2n-1}}
\hat u^{(2n-2)}_{\text{o},i}(\ve k/\epsilon)\right]\,,\label{eq:tnlimit2}
\end{align}
\end{subequations}
Substituting $n=1$ into \eqref{eq:tnlimit1} gives:
\begin{equation} 
\mathcal{\hat T}_i^{(1)}(\ve k) =  
\lim_{\epsilon\to 0} \frac{1}{\epsilon^{2}}
\hat u^{(1)}_{\text{o},i}(\ve k/\epsilon)\,.
\end{equation}
This limit {evaluates to} zero, and we conclude that 
$\mathcal{T}_i^{(1)}(\ve r)\equiv 0$ [Eq.~(\ref{eq:T1})]. Now consider $\mathcal{\hat T}_i^{(3)}(\ve k)$. {First, we use the fact that $\mathcal{\hat T}_i^{(1)}(\ve k)=0$ together with Eq.~\eqref{eq:u1ni} to find the relation}
\begin{align}\label{eq:redord}
        \uouth{2}{i}(\ve k) = \epsilon^{-1} \, \uouth{1}{i}(\ve k)\,.
\end{align}
{It follows from Eqs.~(\ref{eq:tnlimit}) and \eqref{eq:redord} that:
\begin{subequations}
\begin{align}
\mathcal{\hat T}^{(3)}_i(\ve k)
&= \lim_{\epsilon\to 0} \frac{1}{\epsilon^{4}}
\hat u^{(3)}_{\text{o},i}(\ve k/\epsilon)\,,	\\
\mathcal{\hat T}^{(3)}_i(\ve k)
&= \lim_{\epsilon\to 0} \frac{\rm d}{\rm d \epsilon}\left[\frac{1}{\epsilon^{3}}
\hat u^{(1)}_{\text{o},i}(\ve k/\epsilon)\right]\,,	\label{eq:T3lim}
\end{align}
\end{subequations}
Finally, applying Eqs.~\eqref{eq:deltaform2} and \eqref{eq:deltaform3}} we find
\begin{align}
\label{eq:T2sing}
        \mathcal{\hat T}_i^{(3)}(\ve k) &=      
\mathcal{A}_{ij}\frac{\partial}{\partial k_j} \delta(\ve k)\,,
\end{align}
{together with the two equivalent expressions for the normalisation $\mathcal{A}_{ij}$
\begin{subequations}\label{eq:Anorm}
\begin{align}
        \mathcal{A}_{ij}& =  -\int_{\mathbb{R}^3}\!\!\!\ed^3k\,k_j\,\hat u^{(3)}_{\text{o},i}(\ve k)\,, \label{eq:Anorm1}	\\
        \mathcal{A}_{ij}& =  -\int_{\mathbb{R}^3}\!\!\!\ed^3k\,k_j\,\hat u^{(1)}_{\text{o},i}(\ve k)\,.	\label{eq:Anorm2}
\end{align}
\end{subequations}
}
The Fourier transform of Eq.~(\ref{eq:T2sing}) is Eq.~(\ref{eq:T3}), with
\begin{align}
\label{eq:AAp}
	A'_{ij} = \frac{-{\rm i}}{(2\pi)^{3/2}}\mathcal{A}_{ij}\,.
\end{align}
The prefactor comes from the Fourier
transform (we use the symmetric convention).
The integrals ${\cal A}_{21}$ and ${\cal A}_{12}$ must be evaluated numerically. 
{It is interesting to note} that Eq.~(\ref{eq:Anorm1}) agrees with Eq.~(3.37) in Ref.~\cite{lin1970}  (Lin {\em et al.} use a different convention for the  Fourier transform). This confirms that {our expansion in $\ve k$-space} yields the same result {for a sphere} as that obtained by Lin {\em et al.}, using Saffman's asymptotic matching method.
We could not perform the numerical integrations
of {Eq.~(\ref{eq:Anorm1})} with sufficient accuracy (we estimate the relative error of our numerical integrations to be of the order of ten percent). We therefore suspect that the error in the coefficient computed by Lin {\em et al.} occurred in the numerical integration. 

{On the other hand, substituting Eq.~(\ref{eq:Anorm2}) 
into Eq.~\eqref{eq:theory} and} we find an expression that, {for a spherical particle}, is equivalent to the result obtained by by Stone, Lovalenti \& Brady \cite{stone2016}.
We emphasise that Eq.~(\ref{eq:Anorm2}) expresses the coefficients ${\cal A}_{ij}$ (and thus the angular velocity) in terms of $\hat u_{{\rm o},i}^{(1)}$ [Eq.~(\ref{eq:u1ni})]. It is not necessary to compute $\hat u_{{\rm o},i}^{(3)}$, as suggested by \citet{saffman1965} and \citet{lin1970}, and carried out in Eq.~(\ref{eq:Anorm1}). {This detail makes a distinct difference.} {The 
integrals in Eq.~(\ref{eq:Anorm2}) are much simpler to evaluate numerically.}
We find: 
\begin{equation}\label{eq:Aresult}
{\cal A}_{12} = {\rm i}\,0.517\quad \mbox{and}\quad {\cal A}_{21} ={\rm i}\,2.220\,,
\end{equation}
Using Eq.~(\ref{eq:AAp}) we obtain the coefficients quoted in Section \ref{sec:av}.

\section{Direct numerical simulations}
\label{sec:dns}
The numerical simulations are performed using the commercial finite-element software package Comsol Multiphysics 5.1.
The simulation domain is a cube with side length $L$. Since $\kappa = 2a/L$, the
side length is $2\kappa^{-1}$ in dimensionless variables. We impose velocity boundary conditions 
in the shear direction, $u_1 = \pm \kappa^{-1}$ at $r_2 = \pm \kappa^{-1}$,
and periodic boundary conditions
in the flow and vorticity directions at $r_{1,3} = \pm \kappa^{-1}$.  On the particle surface no-slip boundary conditions are used.  For given values of $\Reys$ and $\kappa$ we obtain
the steady-state solution of Eq.~(\ref{eq:nse}), together with the corresponding angular velocity $\omega$.

The numerical mesh consists of tetrahedral elements.
The spatial resolution changes, from $\Delta r_{\rm min}$ close to the particle to $\Delta r_{\rm max}$ far from the particle. All simulations are performed using $\Delta r_{\rm min}=0.06$ and $\Delta r_{\rm max}=0.1\kappa^{-1}$. To assess the numerical precision
we also perform simulations with $\Delta r_{\rm min}=0.12$. The difference $|\omega(\Delta r_{\rm min}=0.12)-\omega(\Delta r_{\rm min}=0.06)|$ estimates the absolute numerical error of the angular velocity. 
We estimate this error to be smaller than  $10^{-6}$ in dimensionless units.

We compute  the creeping-flow limit $\omega^{(0)}$ 
of the angular velocity in  the finite system by evaluating the angular velocity for a fluid with zero mass density. We estimate that 
the absolute error of the finite-size corrections is smaller 
than $10^{-6}$ in dimensionless units.
\begin{table}[t]
\caption{\label{tab:1}
Numerical values of the parameter $b(\kappa)$ in Eq.~(\ref{eq:Re2}), obtained by fitting the results of the direct numerical simulations.
For $\kappa = 0.01$ the fit used values of $\Reys$ smaller than $10^{-3}$. For the other values of $\kappa$ the data was fitted up to
$\Reys=2.5\times 10^{-3}$.}
\begin{tabular}{lcccccccc}
\hline\hline
$\kappa$\,\, & \,\,0.01\,\, &\,\,0.025\,\, &\,\,0.03\,\, &\,\,0.04\,\, &\,\,0.05\,\, &\,\,0.06\,\, & \,\,0.08\,\, & \,\,0.1\\\hline
$b(\kappa)$\,\,       & 1.48   &  0.60  &  0.49&  0.35& 0.27 & 0.21 & 0.14   & 0.10     \\
\hline\hline
\end{tabular}
\end{table}

\section{{Discussion and conclusions}}\label{sec:dc}
{We computed the inertial corrections to the angular velocity of a spheroid log-rolling in an unbounded simple shear to order $\epsilon^3=\Reys^{3/2}$. Our main result, Eq.~\eqref{eq:final_result}, is shown as a function of the particle aspect ratio $\lambda$ by the solid red line in Fig.~\ref{fig:3}. We observe excellent agreement with the results of our direct numerical simulations (shown as open symbols).
Furthermore, we see that the inertial correction is largest for the spherical particle. This reflects the fact that the surface area $|\mathscr{S}|$ 
of the particle is largest when $\lambda=1$. The surface area is also
shown in Fig.~\ref{fig:3}, as a black dashed line. 
As $\lambda\to\infty$ the inertial correction vanishes. This is expected because the latter limit corresponds to slender rods, which do not couple back to the fluid through the boundary conditions. For extended bodies, fluid elements near the rotating particle surface are centrifuged away from the rotation axis due to fluid inertia, thereby slowing the particle down. In the slender-disk limit ($\lambda\to 0$) the inertial correction approaches a finite constant $\sim\Reys^{3/2}$.}

The results of the numerical simulations {for a spherical particle} are plotted {as a function of \Reys} in Fig.~\ref{fig:2},
for three different system sizes: $\kappa=2a/L=0.01$, $0.025$, and $0.05$. Shown is
the difference between $\omega$ and the value of the angular velocity in  creeping-flow limit, $\omega^{(0)}$.
{The inset of} Fig.~\ref{fig:2} shows $\omega^{(0)}$ as a function of $\kappa$.  A fit to 
\begin{align}
\label{eq:wRe0}
\omega^{(0)}(\kappa) = -\tfrac{1}{2} + C \kappa^3\,,
\end{align}
gives $C\approx 0.22$.
For $\Reys$ between {$5\times10^{-4}$ and $5\times 10^{-2}$} the numerical data for the largest system agree well with the theoretical prediction (\ref{eq:result}). This conclusion is supported also by the numerical results reported in Ref.~\cite{mik04}.

But at smaller values of the shear Reynolds number the numerical data lies below the theory. 
The reason is that  $L<\ell_{\rm S}$ at very small values of $\Reys$. In this case the outer region does not exist {and} the problem becomes a regular perturbation problem. {We therefore  expect} that the inertial correction is quadratic in this regime:
\begin{equation}
\label{eq:Re2} \omega = \omega^{(0)}(\kappa) + b(\kappa) \Reys^2\quad\mbox{when}\quad
\Reys \ll\kappa^2 \,.
\end{equation}
The thin solid (blue) lines in Fig.~\ref{fig:2} show that this is indeed the case. Fitting Eq.~(\ref{eq:Re2}) to the numerical data we find that the coefficient $b(\kappa)$ increases monotonously as $\kappa$ decreases. The numerical values of $b(\kappa)$ are given in Table~\ref{tab:1}. 

Also shown in Fig.~\ref{fig:2} are experimental results of Poe \& Acrivos \cite{poe1975}  for a sphere in a shear flow
at shear Reynolds numbers of order unity. At these Reynolds numbers the theory {\eqref{eq:final_result} for $\lambda=1$} is no longer valid.  The results of our numerical simulations are slightly below the experimental results. We do not know the reason for this discrepancy, but note
that Poe \& Acrivos \cite{poe1975} comment on the fact that a re-circulation
flow observed in their experiments  at $\Reys\approx 5$ may have caused 
a spurious increase of the angular velocity. 

{In summary, we} observe
good agreement between the numerical simulations and theory 
where expected, and we find reasonable agreement between 
the simulations and the experiment at $\Reys$ of order unity.

{The details of} our matching method differ from the standard method \cite{proudman1957,saffman1965}. We conclude by briefly summarising the advantages of our approach. We have described two different schemes of calculating the third-order contribution to the angular velocity. The first scheme (Section~\ref{sec:singexp1}) expresses the angular velocity in terms of $\hat u^{(3)}_{\text{o},i}({{\ve k}})$ [Eq.~(\ref{eq:u1ni})].  This approach is equivalent to the standard asymptotic matching scheme used by  Lin {\em et al.} \cite{lin1970}. It results in integral expressions that we could not evaluate accurately enough.

The second scheme allows {us} to express the third-order contribution to the angular velocity in terms of $\hat u^{(1)}_{\text{o},i}({{\ve k}})$. This leads to integrals that are much easier to evaluate.  \citet{saffman1965} remarks that the calculation of such terms in this outer expansion is ``a matter of great difficulty''. Our new scheme is therefore of practical importance when performing asymptotic matching, as we avoid computing two orders {in $\epsilon=\sqrt{\Reys}$} altogether.  Our approach shares 
this advantage with an alternative method that is based on the reciprocal theorem \cite{stone2016}, {and leads to identical integral expressions. 
This provides provides an interesting link between the standard asymptotic matching method  and approaches based on the reciprocal theorem.  
But it remains to be investigated} how general this correspondence is.

{We expect that our scheme offers advantages in different problems too. One example is the evaluation of the viscosity and the normal stresses in a dilute
suspension of spheroids. For spherical particles these were calculated in 
Refs.~\cite{lin1970,sub11,stone2016}.}

{In an unbounded shear the log-rolling orbit is stable
for oblate  particles, but unstable for prolate particles. 
Direct numerical simulations \cite{rosen2015d} show 
that the corresponding stability exponent of the log-rolling orbit has a $\Reys^{3/2}$-correction which is of the same
order as the coefficients of the linear contribution. It would therefore be of interest to compute this correction from first principles. 
It is also important to mention that the direct numerical simulations are performed for a finite system, a bounded shear. It would be of great interest to generalise the results obtained here, but also those reported in
Refs.~\cite{einarsson2015a,einarsson2015b,candelier2015}, to finite systems.
The first step is to compute the small-$\Reys^2$ asymptotes in Fig.~\ref{fig:2}.}

\acknowledgments{We thank Howard Stone for discussions, and for making the preprint 
\cite{stone2016} available to us. We thank Kirsten Mehlig for help with the analysis of the numerical data. JM, JE and BM acknowledge  support by Vetenskapsr\aa{}det [grant number 2013-3992], Formas [grant number 2014-585], by the grant \lq Bottlenecks for particle growth in turbulent aerosols\rq{} from the Knut and Alice Wallenberg Foundation, Dnr. KAW 2014.0048,
and by the MPNS COST Action MP1305 \lq Flowing matter\rq{}.
TR and FL acknowledge financial support from the Wallenberg Wood Science Center, Bengt Ingestr\"{o}m's Foundation, and \AA Forsk.  
The computer simulations were performed using resources provided by the Swedish National Infrastructure for Computing at the National Supercomputer Center in Sweden.

\appendix
\section{}
\label{app:delta}
{In this appendix we derive relation \eqref{eq:deltaform3}. To this end we consider the action of $\lim_{\epsilon\to0}\frac{\ed}{\ed\epsilon}
\big[\epsilon^{-3}h(\ve k/\epsilon)\big]$ upon a test function $\phi$:
\begin{align}
&
\lim_{\epsilon\to0}\Big\langle \frac{\ed}{\ed\epsilon}
\big[\epsilon^{-3}h(\ve k/\epsilon)\big],\phi \Big\rangle\\
&\hspace*{5mm}= \lim_{\epsilon\to0}\Big\langle (-1)\epsilon^{-4}
\big[3+  {k_j}  \frac{\partial}{\partial {k_j}}h(\ve k/\epsilon)\big] .\phi \Big\rangle\nonumber
\end{align}
This expression can written in the form of a divergence:
\begin{align}
&
\lim_{\epsilon\to0}\Big\langle \frac{\ed}{\ed\epsilon}
\big[\epsilon^{-3}h(\ve k/\epsilon)\big],\phi \Big\rangle\\
 &\hspace*{5mm}= \lim_{\epsilon\to0}\Big\langle (-1)\epsilon^{-4}
 \frac{\partial}{\partial { k_j}} \big[{ k_j} h(\ve k/\epsilon)\big],\phi \Big\rangle\nonumber\\
 &\hspace*{5mm}= \Big\langle\lim_{\epsilon\to0}\big[ \epsilon^{-3} \left(\frac{ k_j}{\epsilon}  \right) h(\ve k/\epsilon)\big],\frac{\partial\phi}{\partial k_j} \Big\rangle\,.    \nonumber
\end{align}
We use Eq.~(\ref{eq:deltaform1}) to evaluate the last expression further:
\begin{align}
\lim_{\epsilon\to0}\Big\langle \frac{\ed}{\ed\epsilon}
\big[\epsilon^{-3}h(\ve k/\epsilon)\big],\phi \Big\rangle	&= -\mathcal{A}_{ij}\Big\langle \delta(\ve k),\frac{\partial\phi}{\partial k_j} \Big\rangle\\
&\hspace*{-2.5cm}= \mathcal{A}_{ij}\Big\langle \frac{\partial}{\partial k_j}\delta(\ve k),\phi \Big\rangle\,.
\nonumber
\end{align}
In summary we find: 
\begin{align}
        \lim_{\epsilon\to0}\frac{\ed}{\ed\epsilon}
\big[\epsilon^{-3}h(\ve k/\epsilon)\big] &=
\mathcal{A}_{ij} \frac{\partial}{\partial k_j} \delta(\ve k)
\end{align}
with 
\begin{align}
	\mathcal{A}_{ij}& =  -\int_{\mathbb{R}^3}\!\!\!\ed^3k\,k_j\,h(\ve k)\,.
\end{align}
This is Eq.~(\ref{eq:deltaform3}).}


\begin{thebibliography}{17}%
\makeatletter
\providecommand \@ifxundefined [1]{%
 \@ifx{#1\undefined}
}%
\providecommand \@ifnum [1]{%
 \ifnum #1\expandafter \@firstoftwo
 \else \expandafter \@secondoftwo
 \fi
}%
\providecommand \@ifx [1]{%
 \ifx #1\expandafter \@firstoftwo
 \else \expandafter \@secondoftwo
 \fi
}%
\providecommand \natexlab [1]{#1}%
\providecommand \enquote  [1]{``#1''}%
\providecommand \bibnamefont  [1]{#1}%
\providecommand \bibfnamefont [1]{#1}%
\providecommand \citenamefont [1]{#1}%
\providecommand \href@noop [0]{\@secondoftwo}%
\providecommand \href [0]{\begingroup \@sanitize@url \@href}%
\providecommand \@href[1]{\@@startlink{#1}\@@href}%
\providecommand \@@href[1]{\endgroup#1\@@endlink}%
\providecommand \@sanitize@url [0]{\catcode `\\12\catcode `\$12\catcode
  `\&12\catcode `\#12\catcode `\^12\catcode `\_12\catcode `\%12\relax}%
\providecommand \@@startlink[1]{}%
\providecommand \@@endlink[0]{}%
\providecommand \url  [0]{\begingroup\@sanitize@url \@url }%
\providecommand \@url [1]{\endgroup\@href {#1}{\urlprefix }}%
\providecommand \urlprefix  [0]{URL }%
\providecommand \Eprint [0]{\href }%
\providecommand \doibase [0]{http://dx.doi.org/}%
\providecommand \selectlanguage [0]{\@gobble}%
\providecommand \bibinfo  [0]{\@secondoftwo}%
\providecommand \bibfield  [0]{\@secondoftwo}%
\providecommand \translation [1]{[#1]}%
\providecommand \BibitemOpen [0]{}%
\providecommand \bibitemStop [0]{}%
\providecommand \bibitemNoStop [0]{.\EOS\space}%
\providecommand \EOS [0]{\spacefactor3000\relax}%
\providecommand \BibitemShut  [1]{\csname bibitem#1\endcsname}%
\let\auto@bib@innerbib\@empty
\bibitem [{\citenamefont {Einarsson}\ \emph
  {et~al.}(2015{\natexlab{a}})\citenamefont {Einarsson}, \citenamefont
  {Candelier}, \citenamefont {Lundell}, \citenamefont {Angilella},\ and\
  \citenamefont {Mehlig}}]{einarsson2015a}%
  \BibitemOpen
  \bibfield  {author} {\bibinfo {author} {\bibfnamefont {J.}~\bibnamefont
  {Einarsson}}, \bibinfo {author} {\bibfnamefont {F.}~\bibnamefont
  {Candelier}}, \bibinfo {author} {\bibfnamefont {F.}~\bibnamefont {Lundell}},
  \bibinfo {author} {\bibfnamefont {J.R.}\ \bibnamefont {Angilella}}, \ and\
  \bibinfo {author} {\bibfnamefont {B.}~\bibnamefont {Mehlig}},\ }\bibfield
  {title} {\enquote {\bibinfo {title} {Rotation of a spheroid in a simple shear
  at small {R}eynolds number},}\ }\href@noop {} {\bibfield  {journal} {\bibinfo
   {journal} {Phys. Fluids}\ }\textbf {\bibinfo {volume} {27}},\ \bibinfo
  {pages} {063301} (\bibinfo {year} {2015}{\natexlab{a}})}\BibitemShut
  {NoStop}%
\bibitem [{\citenamefont {Jeffery}(1922)}]{jeffery1922}%
  \BibitemOpen
  \bibfield  {author} {\bibinfo {author} {\bibfnamefont {G.~B.}\ \bibnamefont
  {Jeffery}},\ }\bibfield  {title} {\enquote {\bibinfo {title} {The motion of
  ellipsoidal particles immersed in a viscous fluid},}\ }\href {\doibase
  10.1098/rspa.1922.0078} {\bibfield  {journal} {\bibinfo  {journal}
  {Proceedings of the Royal Society of London. Series A}\ }\textbf {\bibinfo
  {volume} {102}},\ \bibinfo {pages} {161--179} (\bibinfo {year}
  {1922})}\BibitemShut {NoStop}%
\bibitem [{\citenamefont {Bluemink}\ \emph {et~al.}(2008)\citenamefont
  {Bluemink}, \citenamefont {Lohse}, \citenamefont {Prosperetti},\ and\
  \citenamefont {van Wijngaarden}}]{blue2008}%
  \BibitemOpen
  \bibfield  {author} {\bibinfo {author} {\bibfnamefont {J.~J.}\ \bibnamefont
  {Bluemink}}, \bibinfo {author} {\bibfnamefont {D.}~\bibnamefont {Lohse}},
  \bibinfo {author} {\bibfnamefont {A.}~\bibnamefont {Prosperetti}}, \ and\
  \bibinfo {author} {\bibfnamefont {L.}~\bibnamefont {van Wijngaarden}},\
  }\bibfield  {title} {\enquote {\bibinfo {title} {A sphere in a uniformly
  rotating or shearing flow},}\ }\href@noop {} {\bibfield  {journal} {\bibinfo
  {journal} {J.Fluid Mech.}\ }\textbf {\bibinfo {volume} {600}},\ \bibinfo
  {pages} {201} (\bibinfo {year} {2008})}\BibitemShut {NoStop}%
\bibitem [{\citenamefont {Lin}\ \emph {et~al.}(1970)\citenamefont {Lin},
  \citenamefont {Peery},\ and\ \citenamefont {Schowalter}}]{lin1970}%
  \BibitemOpen
  \bibfield  {author} {\bibinfo {author} {\bibfnamefont {C.~J.}\ \bibnamefont
  {Lin}}, \bibinfo {author} {\bibfnamefont {J.~H.}\ \bibnamefont {Peery}}, \
  and\ \bibinfo {author} {\bibfnamefont {W.~R.}\ \bibnamefont {Schowalter}},\
  }\bibfield  {title} {\enquote {\bibinfo {title} {Simple shear flow around a
  rigid sphere: inertial effects and suspension rheology},}\ }\href@noop {}
  {\bibfield  {journal} {\bibinfo  {journal} {J. Fluid Mech.}\ }\textbf
  {\bibinfo {volume} {44}},\ \bibinfo {pages} {1} (\bibinfo {year}
  {1970})}\BibitemShut {NoStop}%
\bibitem [{\citenamefont {Stone}\ \emph {et~al.}(2001)\citenamefont {Stone},
  \citenamefont {Brady},\ and\ \citenamefont {Lovalenti}}]{stone2016}%
  \BibitemOpen
  \bibfield  {author} {\bibinfo {author} {\bibfnamefont {H.~A.}\ \bibnamefont
  {Stone}}, \bibinfo {author} {\bibfnamefont {J.~F.}\ \bibnamefont {Brady}}, \
  and\ \bibinfo {author} {\bibfnamefont {P.~M.}\ \bibnamefont {Lovalenti}},\
  }\bibfield  {title} {\enquote {\bibinfo {title} {Inertial effects on the
  rheology of suspensions and on the motion of individual particles},}\
  }\href@noop {} {\  (\bibinfo {year} {2001})},\ \bibinfo {note}
  {unpublished}\BibitemShut {NoStop}%
\bibitem [{\citenamefont {Candelier}\ \emph {et~al.}(2013)\citenamefont
  {Candelier}, \citenamefont {Mehaddi},\ and\ \citenamefont
  {Vauquelin}}]{candelier2013:b}%
  \BibitemOpen
  \bibfield  {author} {\bibinfo {author} {\bibfnamefont {F.}~\bibnamefont
  {Candelier}}, \bibinfo {author} {\bibfnamefont {R.}~\bibnamefont {Mehaddi}},
  \ and\ \bibinfo {author} {\bibfnamefont {O.}~\bibnamefont {Vauquelin}},\
  }\bibfield  {title} {\enquote {\bibinfo {title} {Note on the method of
  matched-asymptotic expansions for determining the force acting on a
  particle},}\ }\href@noop {} {\bibfield  {journal} {\bibinfo  {journal}
  {arXiv:1307.6314}\ } (\bibinfo {year} {2013})}\BibitemShut {NoStop}%
\bibitem [{\citenamefont {Proudman}\ and\ \citenamefont
  {Pearson}(1957)}]{proudman1957}%
  \BibitemOpen
  \bibfield  {author} {\bibinfo {author} {\bibfnamefont {I.}~\bibnamefont
  {Proudman}}\ and\ \bibinfo {author} {\bibfnamefont {J.~R.~A.}\ \bibnamefont
  {Pearson}},\ }\bibfield  {title} {\enquote {\bibinfo {title} {Expansions at
  small {R}eynolds numbers for the flow past a sphere and circular cylinder.}}\
  }\href@noop {} {\bibfield  {journal} {\bibinfo  {journal} {J. Fluid Mech.}\
  }\textbf {\bibinfo {volume} {22}},\ \bibinfo {pages} {385--400} (\bibinfo
  {year} {1957})}\BibitemShut {NoStop}%
\bibitem [{\citenamefont {Saffman}(1965)}]{saffman1965}%
  \BibitemOpen
  \bibfield  {author} {\bibinfo {author} {\bibfnamefont {P.~G.}\ \bibnamefont
  {Saffman}},\ }\bibfield  {title} {\enquote {\bibinfo {title} {The lift on a
  small sphere in a slow shear flow},}\ }\href@noop {} {\bibfield  {journal}
  {\bibinfo  {journal} {J. Fluid Mech.}\ }\textbf {\bibinfo {volume} {22}},\
  \bibinfo {pages} {385--400} (\bibinfo {year} {1965})}\BibitemShut {NoStop}%
\bibitem [{\citenamefont {Candelier}\ \emph {et~al.}(2016)\citenamefont
  {Candelier}, \citenamefont {Einarsson},\ and\ \citenamefont
  {Mehlig}}]{candelier2016}%
  \BibitemOpen
  \bibfield  {author} {\bibinfo {author} {\bibfnamefont {F.}~\bibnamefont
  {Candelier}}, \bibinfo {author} {\bibfnamefont {J.}~\bibnamefont
  {Einarsson}}, \ and\ \bibinfo {author} {\bibfnamefont {B.}~\bibnamefont
  {Mehlig}},\ }\bibfield  {title} {\enquote {\bibinfo {title} {Angular dynamics
  of small particles in turbulence},}\ }\href@noop {} {\  (\bibinfo {year}
  {2016})}\BibitemShut {NoStop}%
\bibitem [{\citenamefont {Kim}\ and\ \citenamefont {Karrila}(1991)}]{kim1991}%
  \BibitemOpen
  \bibfield  {author} {\bibinfo {author} {\bibfnamefont {Sangtae}\ \bibnamefont
  {Kim}}\ and\ \bibinfo {author} {\bibfnamefont {Seppo~J.}\ \bibnamefont
  {Karrila}},\ }\href@noop {} {\emph {\bibinfo {title} {Microhydrodynamics:
  principles and selected applications}}},\ Butterworth-Heinemann series in
  chemical engineering\ (\bibinfo  {publisher} {Butterworth-Heinemann},\
  \bibinfo {address} {Boston},\ \bibinfo {year} {1991})\BibitemShut {NoStop}%
\bibitem [{\citenamefont {Childress}(1964)}]{childress1964}%
  \BibitemOpen
  \bibfield  {author} {\bibinfo {author} {\bibfnamefont {S.}~\bibnamefont
  {Childress}},\ }\bibfield  {title} {\enquote {\bibinfo {title} {The slow
  motion of a sphere in a rotating, viscous fluid},}\ }\href@noop {} {\bibfield
   {journal} {\bibinfo  {journal} {J. Fluid Mech.}\ }\textbf {\bibinfo {volume}
  {20}},\ \bibinfo {pages} {305--314} (\bibinfo {year} {1964})}\BibitemShut
  {NoStop}%
\bibitem [{\citenamefont {Mikulencak}\ and\ \citenamefont
  {Morris}(2004)}]{mik04}%
  \BibitemOpen
  \bibfield  {author} {\bibinfo {author} {\bibfnamefont {D.~R.}\ \bibnamefont
  {Mikulencak}}\ and\ \bibinfo {author} {\bibfnamefont {J.~F.}\ \bibnamefont
  {Morris}},\ }\bibfield  {title} {\enquote {\bibinfo {title} {Stationary shear
  flow around fixed and free bodies at finite {R}eynolds number},}\ }\href@noop
  {} {\bibfield  {journal} {\bibinfo  {journal} {J. Fluid Mech.}\ }\textbf
  {\bibinfo {volume} {520}},\ \bibinfo {pages} {215} (\bibinfo {year}
  {2004})}\BibitemShut {NoStop}%
\bibitem [{\citenamefont {Poe}\ and\ \citenamefont {Acrivos}(1975)}]{poe1975}%
  \BibitemOpen
  \bibfield  {author} {\bibinfo {author} {\bibfnamefont {G.~G.}\ \bibnamefont
  {Poe}}\ and\ \bibinfo {author} {\bibfnamefont {A.}~\bibnamefont {Acrivos}},\
  }\bibfield  {title} {\enquote {\bibinfo {title} {Closed-streamline flows past
  rotating single cylinders and spheres: inertia effects},}\ }\href@noop {}
  {\bibfield  {journal} {\bibinfo  {journal} {J. Fluid Mech.}\ }\textbf
  {\bibinfo {volume} {72}},\ \bibinfo {pages} {605} (\bibinfo {year}
  {1975})}\BibitemShut {NoStop}%
\bibitem [{\citenamefont {G.}\ \emph {et~al.}(2011)\citenamefont {G.},
  \citenamefont {Koch}, \citenamefont {Zhang},\ and\ \citenamefont
  {Yang}}]{sub11}%
  \BibitemOpen
  \bibfield  {author} {\bibinfo {author} {\bibfnamefont {Subramanian}\
  \bibnamefont {G.}}, \bibinfo {author} {\bibfnamefont {D.L.}\ \bibnamefont
  {Koch}}, \bibinfo {author} {\bibfnamefont {J.}~\bibnamefont {Zhang}}, \ and\
  \bibinfo {author} {\bibfnamefont {C.}~\bibnamefont {Yang}},\ }\bibfield
  {title} {\enquote {\bibinfo {title} {The influence of the inertially
  dominated outer region on the rheology of a dilute dispersion of
  low-reynolds-number drops or rigid particles},}\ }\href@noop {} {\bibfield
  {journal} {\bibinfo  {journal} {J. Fluid Mech.}\ }\textbf {\bibinfo {volume}
  {674}},\ \bibinfo {pages} {307--358} (\bibinfo {year} {2011})}\BibitemShut
  {NoStop}%
\bibitem [{\citenamefont {Ros\'{e}n}\ \emph {et~al.}(2015)\citenamefont
  {Ros\'{e}n}, \citenamefont {Einarsson}, \citenamefont {Nordmark},
  \citenamefont {Aidun}, \citenamefont {Lundell},\ and\ \citenamefont
  {Mehlig}}]{rosen2015d}%
  \BibitemOpen
  \bibfield  {author} {\bibinfo {author} {\bibfnamefont {T.}~\bibnamefont
  {Ros\'{e}n}}, \bibinfo {author} {\bibfnamefont {J.}~\bibnamefont
  {Einarsson}}, \bibinfo {author} {\bibfnamefont {A.}~\bibnamefont {Nordmark}},
  \bibinfo {author} {\bibfnamefont {C.~K.}\ \bibnamefont {Aidun}}, \bibinfo
  {author} {\bibfnamefont {F.}~\bibnamefont {Lundell}}, \ and\ \bibinfo
  {author} {\bibfnamefont {B.}~\bibnamefont {Mehlig}},\ }\bibfield  {title}
  {\enquote {\bibinfo {title} {Numerical analysis of the angular motion of a
  neutrally buoyant spheroid in shear flow at small {R}eynolds numbers},}\
  }\href@noop {} {\bibfield  {journal} {\bibinfo  {journal} {Phys. Rev. E}\
  }\textbf {\bibinfo {volume} {92}} (\bibinfo {year} {2015})},\ \bibinfo {note}
  {063022}\BibitemShut {NoStop}%
\bibitem [{\citenamefont {Einarsson}\ \emph
  {et~al.}(2015{\natexlab{b}})\citenamefont {Einarsson}, \citenamefont
  {Candelier}, \citenamefont {Lundell}, \citenamefont {Angilella},\ and\
  \citenamefont {Mehlig}}]{einarsson2015b}%
  \BibitemOpen
  \bibfield  {author} {\bibinfo {author} {\bibfnamefont {J.}~\bibnamefont
  {Einarsson}}, \bibinfo {author} {\bibfnamefont {F.}~\bibnamefont
  {Candelier}}, \bibinfo {author} {\bibfnamefont {F.}~\bibnamefont {Lundell}},
  \bibinfo {author} {\bibfnamefont {J.R.}\ \bibnamefont {Angilella}}, \ and\
  \bibinfo {author} {\bibfnamefont {B.}~\bibnamefont {Mehlig}},\ }\bibfield
  {title} {\enquote {\bibinfo {title} {Effect of weak fluid inertia upon
  {Jeffery} orbits},}\ }\href@noop {} {\bibfield  {journal} {\bibinfo
  {journal} {Phys. Rev. E}\ }\textbf {\bibinfo {volume} {91}},\ \bibinfo
  {pages} {041002(R)} (\bibinfo {year} {2015}{\natexlab{b}})}\BibitemShut
  {NoStop}%
\bibitem [{\citenamefont {Candelier}\ \emph {et~al.}(2015)\citenamefont
  {Candelier}, \citenamefont {Einarsson}, \citenamefont {Lundell},
  \citenamefont {Mehlig},\ and\ \citenamefont {Angilella}}]{candelier2015}%
  \BibitemOpen
  \bibfield  {author} {\bibinfo {author} {\bibfnamefont {F.}~\bibnamefont
  {Candelier}}, \bibinfo {author} {\bibfnamefont {J.}~\bibnamefont
  {Einarsson}}, \bibinfo {author} {\bibfnamefont {F.}~\bibnamefont {Lundell}},
  \bibinfo {author} {\bibfnamefont {B.}~\bibnamefont {Mehlig}}, \ and\ \bibinfo
  {author} {\bibfnamefont {J.R.}\ \bibnamefont {Angilella}},\ }\bibfield
  {title} {\enquote {\bibinfo {title} {The role of inertia for the rotation of
  a nearly spherical particle in a general linear flow},}\ }\href@noop {}
  {\bibfield  {journal} {\bibinfo  {journal} {Phys. Rev. E}\ }\textbf {\bibinfo
  {volume} {91}},\ \bibinfo {pages} {053023} (\bibinfo {year}
  {2015})}\BibitemShut {NoStop}%
\end{thebibliography}%
\end{document}